\documentclass[10pt,draft]{mytrans} 

\usepackage{amsmath,amsfonts,amssymb}
\usepackage{psfig}
\usepackage{pst-plot}
\usepackage{latexsym}
\usepackage{hhline,eufrak,euscript,delarray}

\usepackage[lined,algonl,boxed]{algorithm2e}

\newtheorem{theorem}{Theorem}

\def\beginremark#1{\smallskip \noindent \textbf{\textit{Remark #1.}}}
\def\endremark{\endtrivlist\unskip}

\def\G{{\mathbb{G}}}
\def\DIAM{D}
\def\SIZE{|X|}
\def\BULLET{\large{$\bullet$}\normalsize}

\def \beq{\begin{equation}}
\def \eeq{\end{equation}}
\def \be{\begin{eqnarray*}}
\def \ee{\end{eqnarray*}}
\def \ben{\begin{eqnarray}}
\def \een{\end{eqnarray}}

\def \L{\left}
\def \R{\right}
\def \B{\Big}

\def\ten{\rightarrow}

\def\proto#1{\textbf{{\texttt{#1}}}}

\begin{document}
\title{Randomized Initialization of a Wireless Multihop Network}
\author{Vlady Ravelomanana\thanks{Vlady Ravelomanana is with the
LIPN -- UMR 7030 (CNRS), Institut Galil\'ee
 Universit\'e de Paris 13, France. 
 E-mail~: vlad@lipn.univ-paris13.fr}}
\maketitle

\noindent
\begin{abstract}
Address autoconfiguration is an important mechanism required 
to set the IP address of a node automatically in a wireless network. 
The address autoconfiguration, also known as initialization or
 naming, consists to give a unique identifier ranging from 
$1$ to $n$ for a set of $n$ indistinguishable nodes.
We consider a wireless network where
 $n$ nodes (processors) are randomly thrown in a square $X$, 
uniformly and independently. We assume that
 the network is synchronous and two 
nodes are able to communicate if they are within distance at most
of $r$ of each other ($r$ is the transmitting/receiving range). 
The model of this paper concerns nodes without the collision detection 
ability: if two or more
 neighbors of a processor $u$ transmit concurrently at the same time,
 then $u$ would not receive either messages.
We suppose also that nodes  know neither the topology of the network nor 
 the number of nodes in the network. Moreover, they start indistinguishable,
 anonymous and unnamed. 

Under this extremal
 scenario, we design and analyze a fully distributed protocol
 to achieve the initialization 
  task for a wireless multihop network
 of $n$ nodes uniformly scattered in a square $X$. We show how
the transmitting range of the deployed stations can affect the 
typical characteristics such as the degrees and the diameter of the network. By allowing the nodes
to transmit at a range $r= \sqrt{\frac{(1+\ell) \, \ln{n} \, \SIZE}{\pi \, n}}$ (slightly
greater than the one required to have a connected network), we 
show how to design a randomized protocol running in expected time
 $O(n^{3/2} \log^2{n})$ in order to assign
 a unique number ranging from $1$ to $n$
 to each of the $n$ participating nodes. 
\end{abstract}

\begin{keywords}
Multihop networks; address autoconfiguration;
self-configuration in ad hoc networks;
randomized distributed protocols; initialization; naming; 
fundamental limits of random networks.
\end{keywords}


\baselineskip=16pt  

\section{Introduction}
Distributed, multihop wireless networks, such as ad hoc networks,
 sensor networks or radio networks, are gaining in importance
as subject of research \cite{PERKINS}. Here, a network is a collection
of transmitter-receiver devices, referred to 
as \textit{nodes} (\textit{stations} or \textit{processors}).

Wireless multihop networks are formed by a group of nodes that
can communicate with each other over a wireless channel. Nodes or processors
come without ready-made links and without centralized controller.
The network formed by these processors 
can be modeled by its \textit{reachability graph}
in which the existence of a directed edge $u \rightarrow v$ means
that $v$ can be reached from $u$. If the power of all
 transmitters/receivers is the same, the underlying reachability graph
is symmetric. As opposed to traditional networks, wireless networks are  often composed
 of nodes whose number can be several orders of magnitude higher than
the nodes in conventional networks \cite{SURVEY}. Sensor nodes are often deployed
 inside a phenomenon. Therefore, the positions of these nodes need not
be engineered or pre-determined. This allows random and rapid deployment in
inaccessible terrains and suit well the specific needs to disaster-relief,
law enforcement, collaborative computing and other 
special purpose applications. 

As customary, the 
time is assumed to be slotted and nodes can send messages in 
synchronous \textit{rounds} or \textit{time slots}. In each round,
every node can act either as a \textit{transmitter} or as \textit{receiver}.
A node $u$ acting as receiver in a given round gets a message, if and only if,
exactly one of its neighbors transmits in the same round. If more than two
neighbors of $u$ transmit simultaneously, $u$ receive nothing. That is,
 the considered networks do not have the ability to distinguish between
absence of message and collision or conflict. 
This assumption is motivated by the fact that
in many real-life situations, the (small) devices in used 
do not always have the collision detection ability. Moreover,
even if such detection mechanism is present, it may be of limited value
 especially in the presence of some noisy channels.
Therefore, it
 is highly desirable to design protocols working independently
of the existence/absence of any collision detection mechanisms.

We consider that a set of  $n$ nodes
are initially \textit{homogeneously scattered} in a square $X$ of size
$\SIZE$. As in several applications, the users of the network can move, and
therefore the topology is unstable. For this reason, it is desirable
for the protocols to refrain from assumptions about the network
topology, or about the information that processors have concerning
the topology. In this work, we assume that none of the processors
have initially any topological information, except the measure (surface)
$\SIZE$ of the square $X$ where they are randomly dropped. We pinpoint
here that even if $\SIZE$ is exactly known but not $n$ then
 even if $n=O(\SIZE)$, equation such as (\ref{DILBERT}) in the theorem \ref{DEGREE} (see below)
 allows us to handle the subtle changements involved in the
  constant hidden by the ``big-Oh'' between $n$ and $O(\SIZE)$.

\textit{Self-configurations} of
 networking devices appear to be one of the most important challenges
in wireless and mobile computing. 
Before networking, each node must have a
\textit{unique identifier} (referred to as \textit{ID} or \textit{address}) 
 and it is highly desirable to have self-configuration protocols for the nodes.
 A mechanism that allows the network
 to create a unique address (ID) automatically for each of 
 its participating nodes is known as the \textit{address autoconfiguration} 
 protocol. In this work, our nodes 
 are initially \textit{indistinguishable}. This assumption arises 
 naturally since it may be either difficult or impossible to get
 interface serial numbers while on missions.  Thus,
 the IDs self-configuration protocols do not have to rely 
 on the existence of serial numbers. 

The problem we address here is then to design
 a \textit{fully distributed protocol} for the address 
autoconfiguration problem (also known as \textit{initialization} \cite{OLARIU}
 or \textit{naming} problem). By distributed protocol, we mean
without the need of any preexisting centralized controller or base station,
 or requiring human interventions (network administrators).

To this end, we remark first that the transmitting range of 
each station can be set to some 
value $r$ ranging from $0$ to $r_{\textsc{max}}$. 
This model is commonly used in mobile computing and radio
networking \cite{CHENG,JUNG-NITIN,MEGUERDICHIAN1,MEGUERDICHIAN2}.
Note that such model is frequently encountered in many domains 
ranging from statistical physics to epidemiology (see for example 
\cite{HALL} for the theory of 
coverage processes or \cite{MEESTER-ROY} for percolative ingredients). 
The random graphs generated this way 
have been considered first in the seminal paper of Gilbert \cite{GILBERT} 
(almost at the same time Erd\"{o}s and R\'enyi considered the
$\G(n,p)$  model \cite{ER60}) and 
analysis of their
properties such as connectivity and coverage have been the subject of intense
 studies \cite{Kumar-Gupta,MILES,PENROSE2,PENROSE3,PENROSE,PENROSE-RGG}. 
It is easy to see that if the transmitting range of the devices is set
to $0$, the underlying graph has no edges. 
If the transmitting range is too large, the graph is
extremely dense making the scheduling of communications difficult. The
Figure \ref{RANDOM_DISK_GRAPH} below shows devices which have been
deployed on some field in a random fashion. 
 The examples of the figure suggest that
 transmission ranges can play a crucial role when
setting protocols at least for randomly distributed nodes. Other 
important parameters are
 the number $n$ of active stations, the shape of the area $X$
where the nodes are scattered and the nature of
 the communications to be established.
For instance,  in \cite{Kumar-Gupta}
the authors considered a set of $n$ nodes and
a disk of unit area. In this case, if the  range of transmission 
of the stations is set to a value $r = r(n)$ satisfying 
$ \pi \, r(n)^2 =  (\log n + \omega(n))/n$, 
it was shown by Gupta and Kumar
that the wireless network is \textit{asymptotically almost surely}
 (a.a.s., for short) connected if and only if $\omega(n)$ tends to $\infty$ 
with $n$. Throughout this paper, an 
event ${\mathcal{E}}_n$ is said to occur
 asymptotically almost surely  if and
only if  the probability 
 $Pr\left[ {\mathcal{E}}_n \right]$
 tends to $1$ as $n \ten \infty$. 
 We also say ${\mathcal{E}}_n$ occurs 
 \textit{with high probability} (w.h.p. for short).

\vspace{-0.0cm}
 \begin{center}
   \begin{figure}[h]
     \begin{minipage}[t]{3.0cm}
       \begin{center}
	 \psfig{file=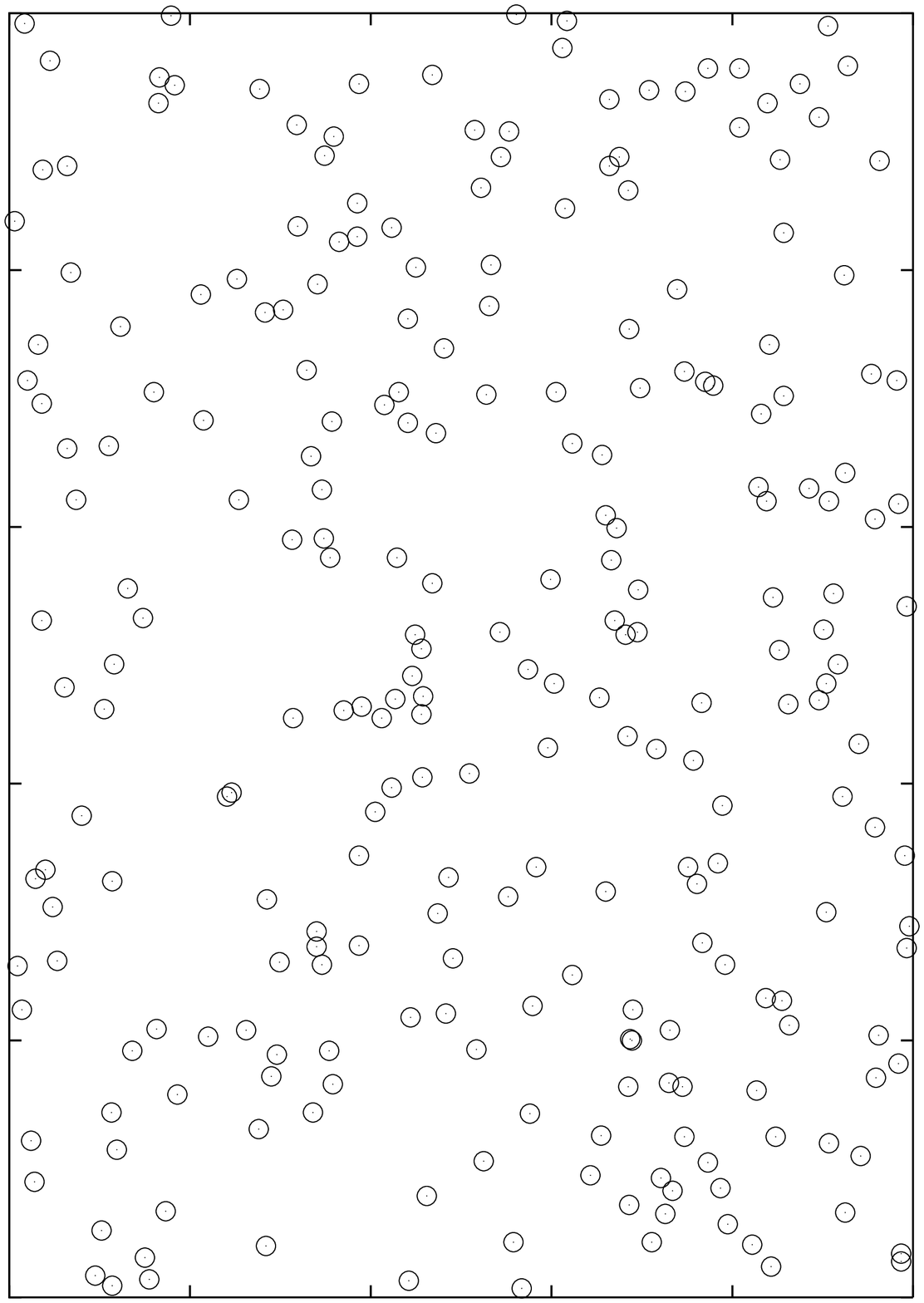,width=3.0cm,height=3.0cm,angle=0}
       \end{center}
       \label{FIG:POISSON1} 
     \end{minipage}
     \hfill
     \begin{minipage}[t]{3.0cm}
       \begin{center}
	 \psfig{file=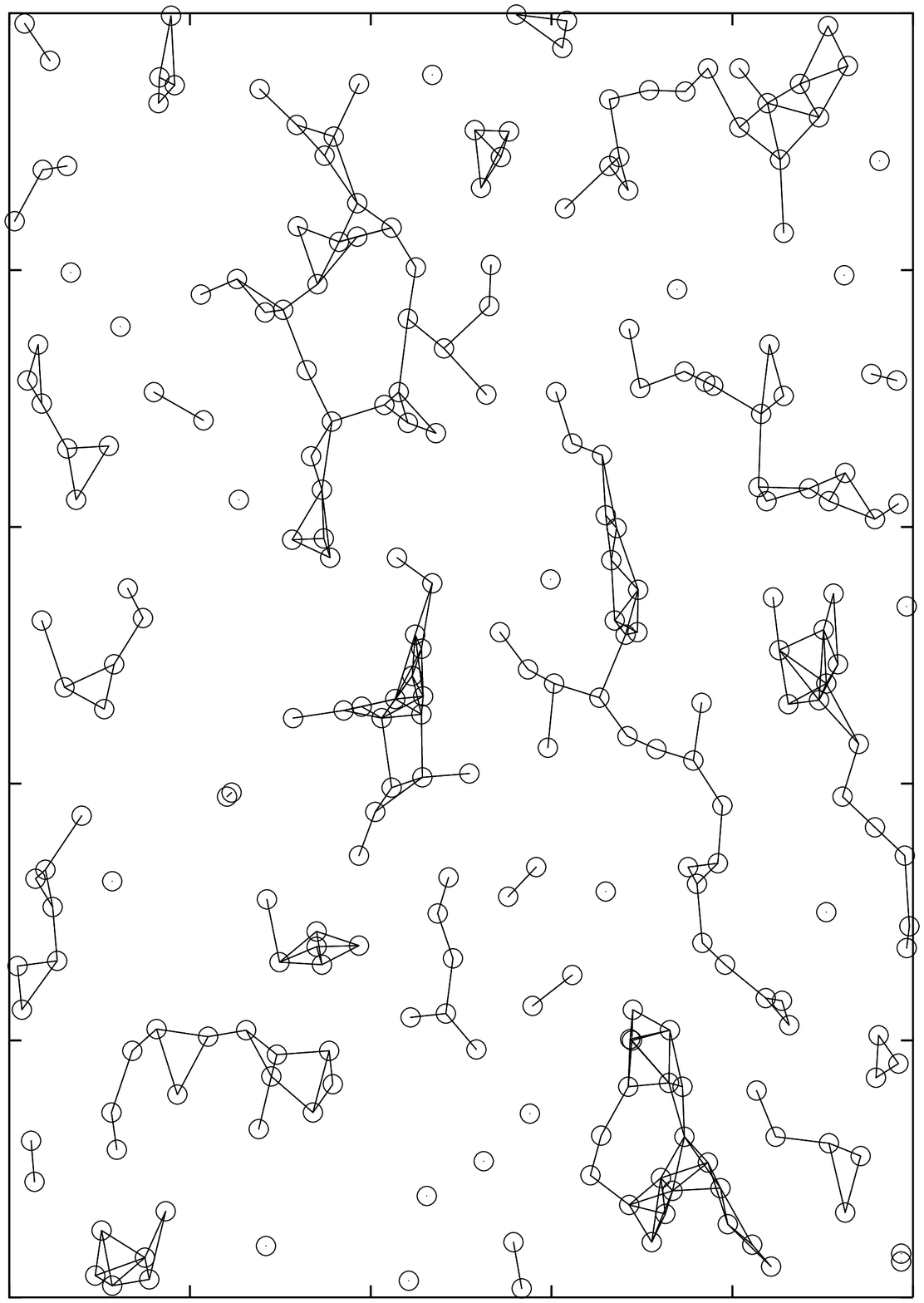,width=3.0cm,height=3.0cm,angle=0}
       \end{center}
       \label{FIG:POISSON2} 
     \end{minipage}
     \hfill
     \begin{minipage}[t]{3.0cm}
       \begin{center}
	 \psfig{file=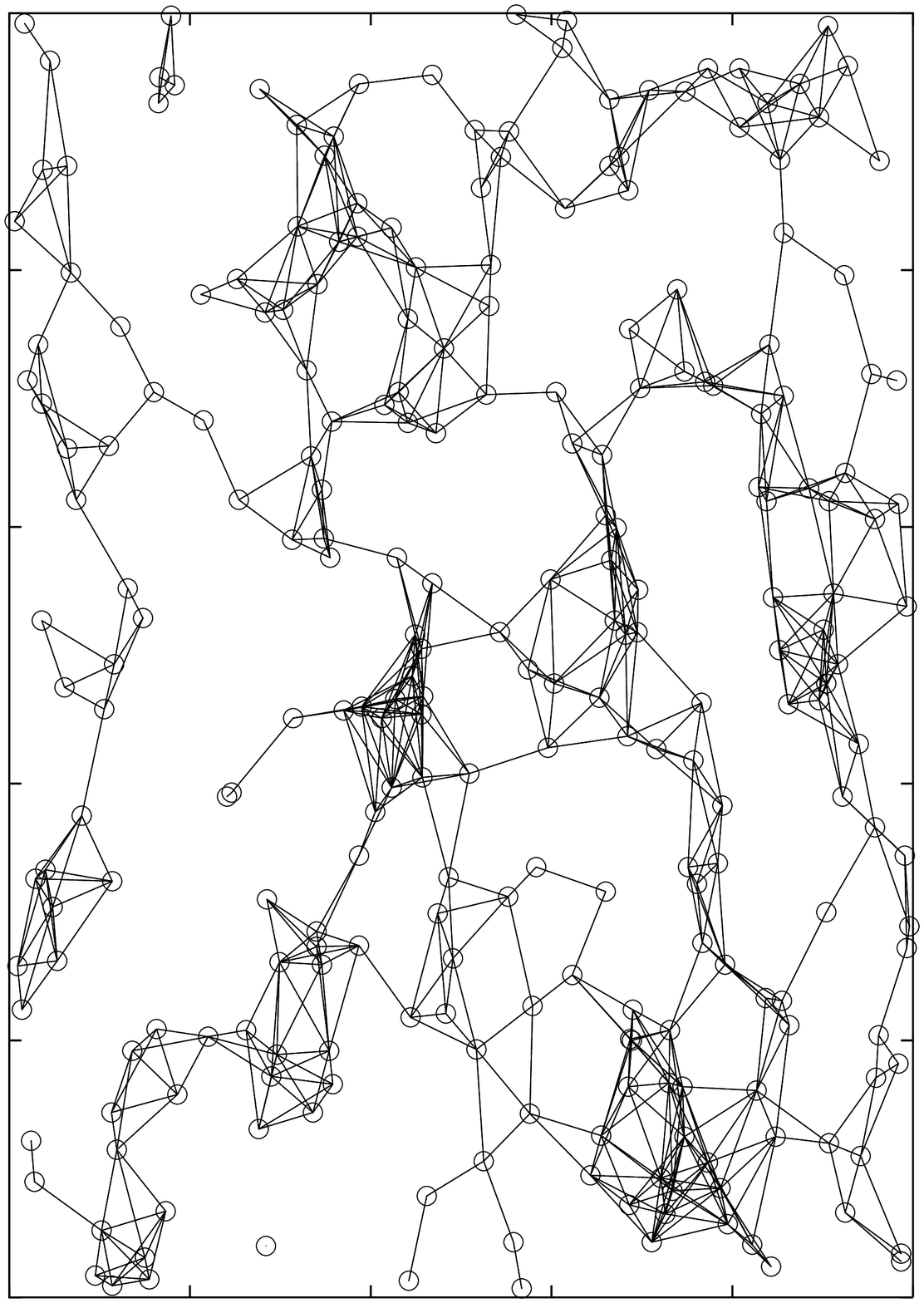,width=3.0cm,height=3.0cm,angle=0}
       \end{center}
       \label{FIG:POISSON4} 
     \end{minipage}
     \hfill
     \begin{minipage}[t]{3.0cm}
       \begin{center}
	 \psfig{file=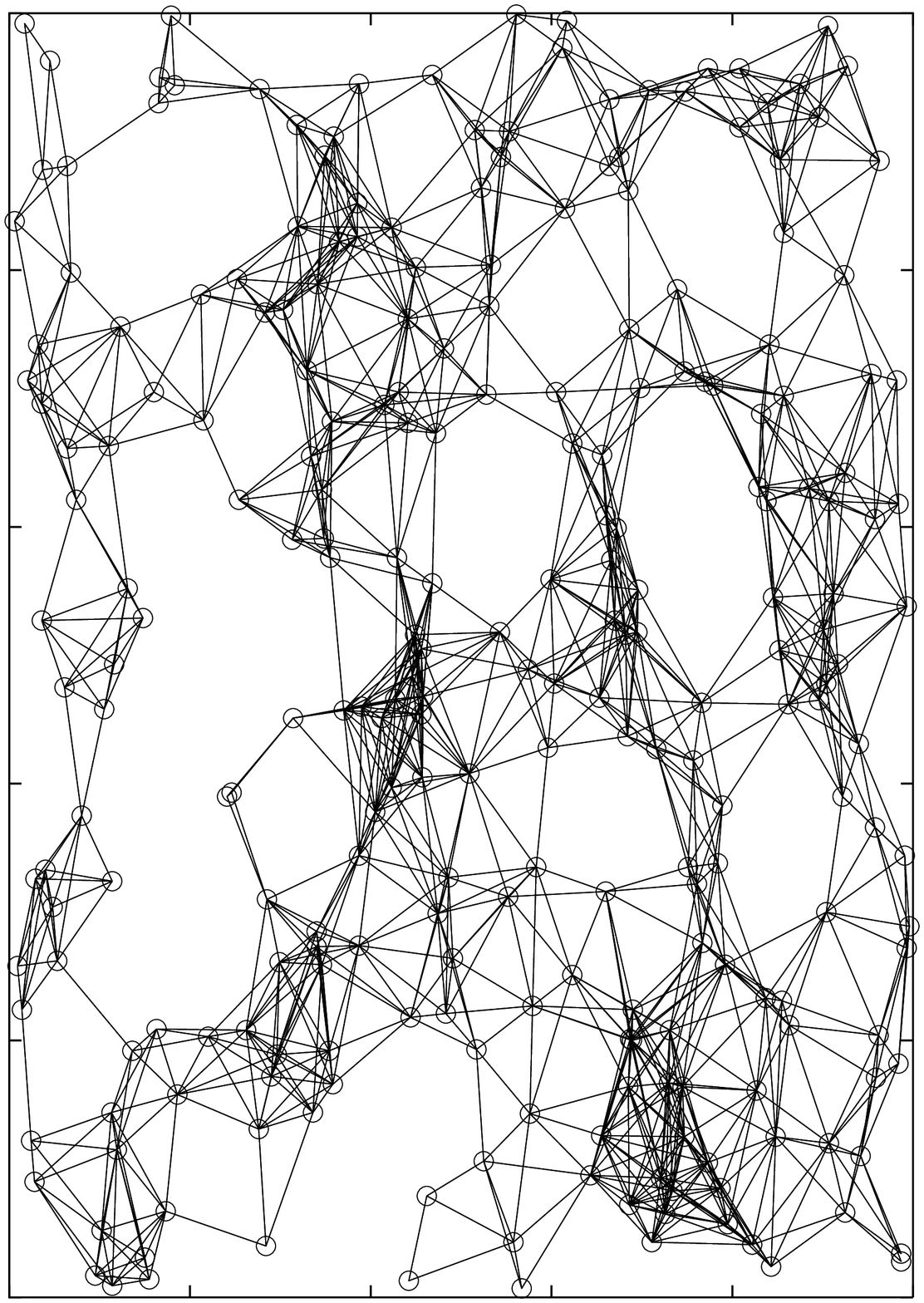,width=3.0cm,height=3.0cm,angle=0}
       \end{center}
     \end{minipage}
     \hfill
     \begin{minipage}[t]{3.0cm}
       \begin{center}
	 \psfig{file=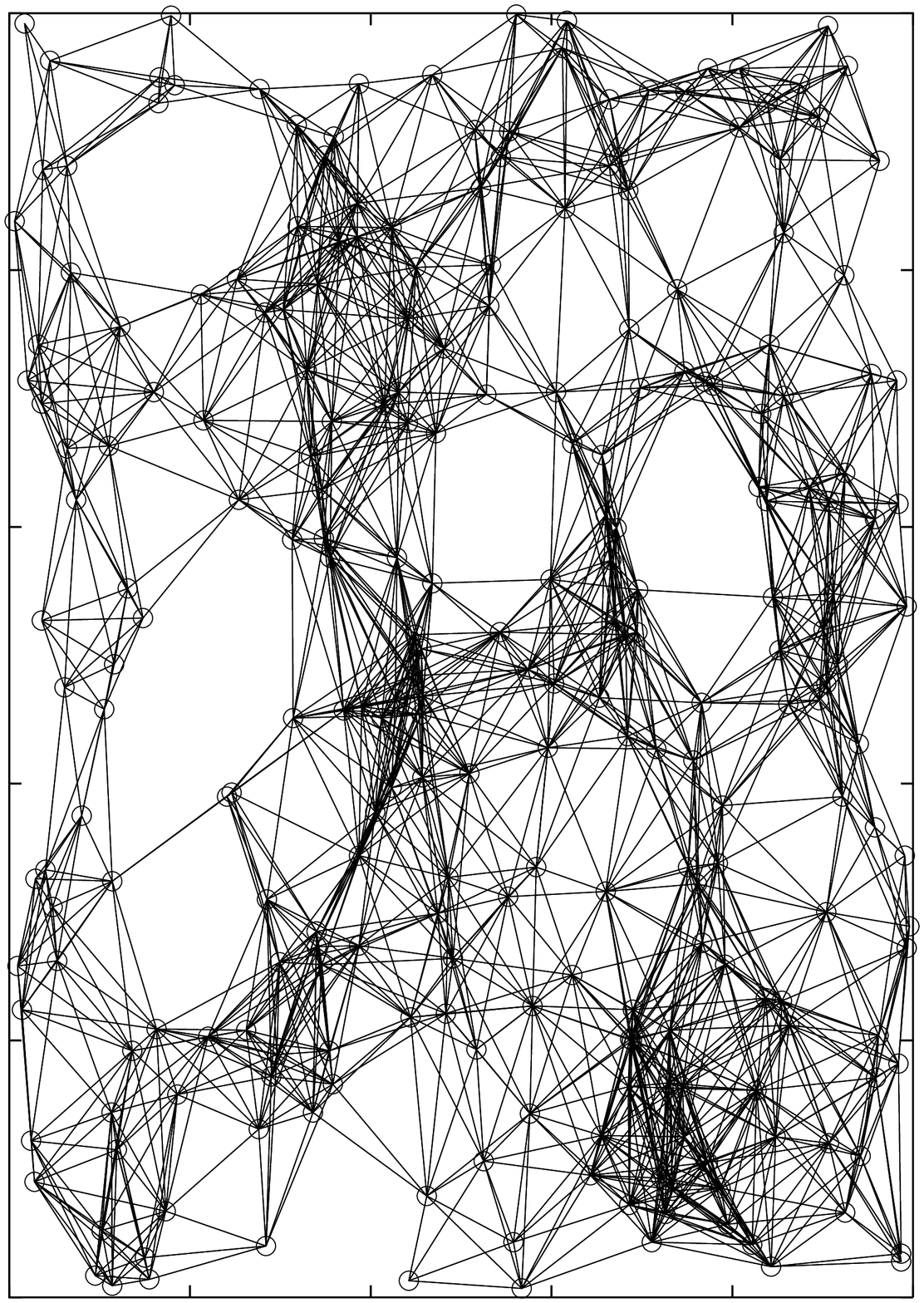,width=3.0cm,height=3.0cm,angle=0}
       \end{center}
     \end{minipage}
     \vspace{-0.5cm}
     \caption[]
     {A typical radio network is generated via uniform distribution of the
$x$ and $y$ coordinates of the devices and the  transmission ranges
of the nodes are increased gradually (from left to right). 
The two last pictures show that if the obtained graph has more 
edges than needed, the number of colliding packets is 
more difficult to control.}
     \label{RANDOM_DISK_GRAPH}
   \end{figure}
 \end{center}
According to these observations,
 to design efficient protocols, we have to take into account and
to exploit 
the structural properties 
 of the reachability graph. In our scenario, since
 none of the nodes knows the number $n$ of the processors in the network,
 our first task is to find distributed protocols 
 that allow (probabilistic) counting of these nodes. We then go on to
 show that by setting the transmitting range parameter correctly,
 the network can be auto-initialized in expected average
 $O(n^{3/2} \ln^2{n})$ time slots. As far as we know,
this is the first analysis for the initialization protocols in the multihop cases
(the single-hop cases have been treated in the litterature in \cite{HAYASHI,OLARIU,OLARIU2,OUR-PAPER}).
Our algorithms are shown to
 take advantage of the fundamental characteristics
 of the network. These limits are computed with the 
help of fully distributed protocols and once known, 
  a divide-and-conquer algorithm is run to assign 
 each of the $n$ processors a distinct ID number in the range
from $1$ to $n$. Even though the protocols are probabilistic, 
once the IDs are attributed their uniqueness can be checked deterministically
by for example the use of deterministic algorithms such as
those of Chrobak, Gasieniec and Rytter in \cite{CHROBAK}.
  As a result, the combination of both protocols
 leads to an initialization protocol  which \textit{always succeeds} and only whose running time
 is random.

Under the conditions described above,
 the Figures 2 and 3 summarize briefly
 the input and output of the distributed 
protocols presented in this work.

\vspace{-2.75cm}
\begin{figure}[h]
\hfill
  \begin{minipage}[t]{7.5cm} 
    \begin{center} 
      \begin{center}
      \begin{pspicture}(0,0)(6,6)
	\psset{unit=0.60}
	\rput(2.2,5.5){\textsc{Input:}}
	\pspolygon (5,0)(5,5)(0,5)(0,0)(5,0) 	
	\rput(1.2,2.3) { $\mathbf{\bullet}$}
	\rput(0.5,1.6) { $\mathbf{\bullet}$}
	\rput(3.2,1.3) { $\mathbf{\bullet}$}
	\rput(4.4,4.3) { $\mathbf{\bullet}$}
	\rput(4.2,0.6) { $\mathbf{\bullet}$}
	\rput(3.0,0.5) { $\mathbf{\bullet}$}
	\rput(2.8,2.2) { $\mathbf{\bullet}$}
	\rput(4.1,2.6) { $\mathbf{\bullet}$}
	\rput(2.6,4.1) { $\mathbf{\bullet}$}
	\rput(3.5,3.5) { $\mathbf{\bullet}$}
	\rput(1.2,4.3) { $\mathbf{\bullet}$}
	\rput(1.1,3.6) { $\mathbf{\bullet}$}
	\rput(0.1,2.3) { $\mathbf{\bullet}$}
	\rput(2.0,2.9) { $\mathbf{\bullet}$}
	\rput(4.2,1.3) { $\mathbf{\bullet}$}
	\rput(3.0,3.9) { $\mathbf{\bullet}$}
	\rput(1.8,4.8) { $\mathbf{\bullet}$}
	\rput(1.1,1.6) { $\mathbf{\bullet}$}
	\rput(0.1,4.3) { $\mathbf{\bullet}$}
	\rput(0.9,2.8) { $\mathbf{\bullet}$}
	\rput(0.2,3.3) { $\mathbf{\bullet}$}
	\rput(1.5,1.0) { $\mathbf{\bullet}$}
	\rput(0.5,0.3) { $\mathbf{\bullet}$}
	\rput(2.2,0.7) { $\mathbf{\bullet}$}
      \end{pspicture}
       \end{center} 
	\caption{$n$ indistinguishable and identical processors
	  randomly placed in the square $X$. The only knowledge
         recquired is the size $\SIZE$ of the support.}
	\label{FIG:INPUT}
    \end{center}
  \end{minipage}
  \hfill
  \begin{minipage}[t]{7.5cm} 
    \begin{center}	
      \begin{pspicture}(0,0)(6,6)
	\psset{unit=0.60}
	\rput(2.2,5.5){\textsc{Output:}}
	\pspolygon (5,0)(5,5)(0,5)(0,0)(5,0) 	
	\rput(1.2,2.3){\small 1}
	\rput(0.5,1.6) {  \small 2}
	\rput(3.2,1.3) {  \small 3}
	\rput(4.4,4.3) {  \small 4}
	\rput(4.2,0.6) {  \small 5}
	\rput(3.0,0.5) {  \small 6}
	\rput(2.8,2.2) {  \small 7}
	\rput(4.1,2.6) {  \small 8}
	\rput(2.6,4.1) {  \small 9}
	\rput(3.5,3.5) {  \small 10}
	\rput(1.2,4.3) {  \small 11}
	\rput(1.1,3.6) {  \small 12}
	\rput(0.1,2.3) {  \small 13}
	\rput(2.0,2.9) {  \small 14}
	\rput(4.2,1.3) {  \small 15}
	\rput(3.0,3.9) {  \small 16}
	\rput(1.8,4.8) {  \small 17}
	\rput(1.1,1.6) {  \small 18}
	\rput(0.1,4.3) {  \small 19}
	\rput(0.9,2.8) {  \small 20}
	\rput(0.2,3.3) {  \small 21}
	\rput(1.5,1.0) {  \small 22}
	\rput(0.5,0.3) {  \small 23}
	\rput(2.2,0.7) {  \small 24}
      \end{pspicture}
      \caption{Each of the $n$ processors is assigned
	a unique ID ranging from $1$ to $n$. These IDs can serve
        as IP address.}
      \label{FIG:OUTPUT}
    \end{center}
  \end{minipage}
  \hfill
\end{figure}
\vspace{0.1cm}

The remainder of this paper is organized as follows. Section 2 
 first presents a randomized protocol \proto{SEND} 
which plays a central key role throughout this paper. We then analyze
 this protocol. In Section 3, we discuss about how to
 set correctly the transmission range of the nodes. Section 3 also offers
 results about the relationship between the transmission range $r$, 
the number of active nodes $n$, the size of $X$,
the maximum degree $\Delta$ and the hop-diameter $\DIAM$ 
of the wireless networks. These results and
the use of the procedure \proto{SEND} allow us 
to build a \proto{BROADCAST} protocol. Section 3 ends with the
design and analysis of a protocol called \proto{SFR}
which plays a central key role  to check the 
correct transmission range.
In Section 4, we briefly recall how to initialize $n$ processors
in the case of single-hop network (i.e., whenever 
the underlying reachability graph
is complete). Using the \proto{BROADCAST} protocol, we then turn on the
design and analysis of a randomized address autoconfiguration 
protocol in the case of wireless multihop networks.

\section{The basic protocol for sending information}
First of all, no deterministic protocol can work correctly in the networks
when processors are anonymous. This can be easily checked: conflict between
 two processors absolutely identical can not be solved deterministically.
Therefore, this impossibility result implies the use of randomness 
(see \cite{EXP-GAP}). Since our processors do not have unique
identifiers, our first task is to build a basic protocol for 
 the nodes which compete to access the unique channel of communication
in order to send a given message. This can be achieved by 
organizing a flipping coin game between them. Recall also that if the
transmission/receiving range is set to a value $r$, only neighbors
of distance at most $r$ are able to communicate when conflicts are
absent. Thus, in the following procedure we have to take into account this
parameter as well as the duration $T$ of the trials:

\baselineskip=12pt

\noindent
\rule{\textwidth}{0.4pt}\\
\noindent {\textbf{Procedure} \proto{SEND}}($msg$, $T$, $r$) \\
\noindent \textbf{For} { $i$ from $0$ to $T$} \textbf{do} \\
\indent \indent  With probability $1/2^i$ send $msg$ to every neighbor \\
\indent \indent  ($\star$ processor within distance at most $r$ $\star$) \\
\noindent  \textbf{end.} \\  
\rule{\textwidth}{0.4pt}

\baselineskip=16pt

\smallskip
\noindent
Note that $r$ is here a parameter which can be tuned to 
a precise value. Again, it should be clear now that only neighbors
within distance at most $r$ can receive the message when
there is no conflict. We have the following result:

\smallskip

\begin{theorem} \label{TH_SEND}
Denote by $r_{\textsc{con}}$ the transmission range required
to get a connected reachability graph.
Let $r \geq r_{\textsc{con}}$ 
be the current transmission range of the processors.
For a fixed node $v$, denote by $d_v$ its degree with $d_v = d_v(r)$.
Suppose that each of the $d_v$ neighbors of $v$ start the execution of
\proto{SEND}$(msg,T,r)$ at the same time. Let $P(T,d_v)$ be the probability
that $v$ will receive the message $msg$ at least once between
the time $t=0$ and $t=T$. Then $P(T,d_v)$ satisfies:\\
\textbf{(i)} the limit $\lim_{T \ten \infty} P(T,d_v)$ exists \\
\textbf{(ii)} if $d_v \ten \infty$ but $2^T \gg d_v$ then
\beq
P(T,d_v) \geq 0.8116... + O\L(\frac{d_v}{2^T}\R) + O\L(\frac{1}{d_v}\R) \, .
\eeq
\end{theorem}

\begin{proof}
The assumption that the reachability graph is connected insures
that for all processor $v$, the degree of $v$ verifies $d_v >0$.
To prove (i), it suffices to observe that $(P(T,d_v))_T$ 
is an increasing sequence bounded by $1$ so that it converges.
For (ii), we have
\beq
P(T, \, d_v) = 1 - \prod_{i=0}^{T} %
\L( 1 - \frac{ {d_v \choose 1} }{2^i} \L(1- \frac{1}{2^i} \R)^{d_v-1} \R) \,.
\label{PTDV}
\eeq
Denote by $s_t(d)$ the quantity:
\ben
s_t(d) = \prod_{i=1}^{t} %
\L(1 - \frac{d}{2^i}\L(1-\frac{1}{2^i}\R)^{d-1} \R).
\een
For any given $i_1$ and for all $i \geq i_1$, we have
\beq
\L(1-\frac{d}{2^i}\B(1-\frac{1}{2^i}\B)^{d} \R) %
\leq \Bigg(1-\frac{d}{2^i} \exp{\L( -\frac{d}{2^i}\B(1+\frac{1}{2^i}\B)\R)} \Bigg) %
\leq  \L(1-\frac{d}{2^i}\exp{\L( -\frac{d}{2^i}\L(1+\frac{1}{2^{i_1}}\R)\R)} \R).
\eeq
Therefore, if  $t \ten \infty$ such that
$2^t \gg d$ by choosing
$i_1 = \lceil \frac{1}{2}\log_2{d} \rceil $ we obtain
\be
s_t(d) & \leq & \prod_{i=i_1}^{t} \Bigg(1-\frac{d}{2^i} %
\exp{\L( -\frac{d}{2^i}\B(1+\frac{1}{2^{i_1}}\B) \R)} \Bigg) \cr
 & \leq & \exp{\L(-\sum_{m\geq 1} \frac{1}{m} \sum_{i=i_1}^{t} %
\frac{d^m}{2^{im}} \exp{\L(-\frac{dm}{2^i}\B(1+\frac{1}{2^{i_1}}\B)\R)}\R)} \cr
& \le & .1884 \;+\; O\L(\frac{1}{d}\R) \;+\; O\L(\frac{d}{2^{t}}\R).
\label{1884}
\ee
We used the so-called Mellin transform asymptotics detailed in
\cite{PHILIPPE-ROBERT} and in \cite[p.~131]{KNUTH}. The value
$\exp{\L(- \sum_{m\geq 1} m!/(m^{m+2}\log{2})\R)} = .188209\ldots$
has been numerically computed with Maple. Note that similar
 computations show that 
$s_t(d) \geq .1883  + O(1/d) + O(d/2^t)$ but
 the above inequality suffices to obtain (\ref{PTDV}).
\end{proof}

\beginremark{1}
Denote by $T^{\star}$ the duration of the coin flipping game
between $d$ nodes needed to succeed to send a message
to a common neighbor. Theorem \ref{TH_SEND} asserts  
that for a suitable value of the transmission range
$r$ such that the graph is connected, $T^{\star} = T^{\star}(d)$ 
is in probability at most $\log_2{d}+O(1)$. That is the probability
$Pr( T^{\star} > \log_2{d} + \omega(d)) \ten 0$ for any arbitrary
function $\omega(d)$ tending to infinity with $d$. This can be
 checked by standard probabilistic arguments 
 since as soon as $2^T \gg d$, the probability of conflict is
less than 0.188... .
\endremark

\smallskip

\beginremark{2}
In \cite{EXP-GAP}, Bar-Yehuda \textit{et al.} have designed a
randomized procedure called \proto{DECAY} to send information. 
They have shown that  if $d$ neighbors
of a node $v$ execute \proto{DECAY} simultaneously, then after 
 $T$ time slots with  $T \geq 2 \log_{2}(d)$ and 
with probability greater than $0.5$, the node $v$ receives a message 
 \cite[pp 108--109]{EXP-GAP}.
In our procedure \proto{SEND}, the proof of 
theorem \ref{TH_SEND} (see also \cite{PHILIPPE-ROBERT}) shows that by changing
the basis of the coin flipping game, viz. replacing
the probability $1/2^i$ in the algorithm by $1/a^i$ for any constant
 $a>1$, the probability of success of the $T$ trials 
 can be made arbitrary close to $1$ (after similar 
logarithmic number of time slots satisfying $a^T \gg d$).
\endremark

\smallskip

In the next Section, we turn on the problem of finding
 suitable values of transmission range whenever the only
\textit{a priori} knowledge of the processors is the size of the 
square $X$. 

\section{Transmission ranges and characteristics of the network}
The aim of this Section is to provide randomized distributed algorithms
that allow the nodes in the network to find 
 the right transmission range such that the reachability
graph is at least connected. To this end, we need to know the relationships
between the transmission range $r$, the number of processors $n$ and
the measure $\SIZE$ of the square. 
Other fundamental characteristics of the graph,
such as the minimum (resp. maximum)
degree $\delta$ (resp. $\Delta$) and the hop-diameter $\DIAM$ 
are also of great interest when designing wireless protocols 
(see \cite{EXP-GAP}). 
Moreover, the limits of the randomly generated network of processors 
help as it will be shown shortly. We refer here to
\cite{GILBERT,Kumar-Gupta,MILES,PENROSE-RGG,BLOUGH,SHAKKOTTAI} for 
works (under various assumptions) related to the fundamental characteristics
and limits of random plane networks. For sake of clarity,
 we treat in this Section two distinct paragraphs:
\begin{itemize}
\item[$\bullet$] The first one concerns the characteristics
of the reachability graph in the superconnectivity regime,
 i.e. when the radius of transmission of the nodes grows much more
faster than the one required to achieve the connectivity of the graph.
\item[$\bullet$] The second subsection is devoted to the design and
analysis of a distributed protocol, called \proto{SFR},
 that will allow the nodes to approximate
their number. At this stage, the technical characteristics
 related to fundamental limits of the graph relating
 $r$, $n$, $\SIZE$, $\DIAM$, $\Delta$ described previously, 
  become extremely important.
\end{itemize}

\subsection{Fundamental limits of a random network in the superconnectivity regime}
\noindent
For several reasons, we follow here the Miles's model \cite{MILES}.
In this model, a large number $n$ of devices
 are dropped in some area $X$. As $n \ten \infty$ but $n = O(\SIZE)$,
the graph generated by the transmitting devices can be well approximated
with a Poisson point process (see for instance \cite{HALL}). First of all,
 this extreme independance property allows penetrating analysis.  Next, since
 Poisson processes are \textit{invariant} if their points are independently
translated (the translations being identically distributed following
some bivariate distribution), the results can take their importance
 for \textit{moving nodes} and therefore, they 
are well suited to cope with randomly deployed mobile devices.
Third, due to Poisson processes properties, if 
with probability $p$, such that $p \times n = O(\SIZE)$, some nodes are
 \textit{faulty} or intentionally \textit{asleep} (e.g. to save
batteries to design energy-efficient algorithms), 
our results remain valid. In this latter scenario,
 the number of nodes $n$ is simply replaced by $n'=p\, n$.

Among other results, Penrose \cite{PENROSE} proved (with our notations)
that if $n/\SIZE = O(1)$ and if $r_{\textsc{con}}$ denotes
the minimal radius of transmission to achieve connectivity then
\beq
\lim_{n \ten \infty} Pr\left[ \pi \, \frac{n}{\SIZE} %
 \, r_{\textsc{con}}^2 - ln(n) %
\leq \omega \right] =  \exp{( -e^{-\omega})} , %
\qquad \omega \in \mathbb{R}\, .
\label{PENROSE_CONNECTIVITY}
\eeq
Penrose's result tells us that by letting
the radius of transmission range growing as 
\beq
r = \sqrt{ \frac{\ln{n} + \omega(n) }{ \pi n} \, \SIZE } \, ,
\eeq
for any arbitrary function of $\omega(n)$ tending to infinity with $n$,
the obtained graph of the network is a.a.s. connected. 

For our purpose, we need the following results related to the degrees 
of the nodes participating in the network according to successive
values of the transmission range:

\smallskip

\begin{theorem} \label{DEGREE}
Denote by $r$ the transmission range of the $n$ nodes randomly distributed
in the square $X$ of size $\SIZE=O(n)$. Then, in the following regimes
with high probability the graph is connected and 
we have:
\begin{itemize}
\item[(i)] For fixed values of $k$, that is $k=O(1)$, if
$\pi \frac{n}{\SIZE} r^2 = \ln{n} + k \ln\ln{n} + \omega(n)$, 
then the graph has a.a.s. a minimum degree of $\delta = k$.
\item[(ii)] Let $k=k(n)$ but $1 \ll k(n) \ll \ln{n}/\ln{\ln{n}}$.
If $\pi \frac{n}{\SIZE} r^2 = \ln{n} + k(n) \, \ln\ln{n}$,
then the minimum degree (resp. maximum degree) is a.a.s. 
$\delta =  k(n)$ (resp. $\Delta = e \, \ln{n}$).
\item[(iii)] If $\pi \frac{n}{\SIZE} r^2 = (1+\ell) \ln{n}$
with $\ell > 0$ then each node $v$ of the graph has
a.a.s. $d_v$ neighbors with
\beq
    - \, \frac{ \ell \,\ln{n}}{W_{-1}\L(- \frac{\ell}{e \, (1+\ell)}\R)} +
o\L( \ln{n} \R) \leq d_v %
\leq  - \, \frac{ \ell \,\ln{n}}{W_{0}\L(- \frac{\ell}{e \, (1+\ell)}\R)} +
o\L( \ln{n} \R) \, ,
\label{DILBERT}
\eeq
where $W_{-1}$ and $W_0$ denote the two branches
 of the Lambert W function\footnote{The Lambert W function
is considered as a special function of mathematics on its own and its
computation has been implemented in mathematical software such as Maple.}
 which are detailed in \cite{LAMBERTW}. See also the appendix of this paper for
some details about the two branches of the Lambert W function.
\end{itemize}
Each geographical point of the support $X$ is also recovered by
$\Theta(\ln{n})$ disks of transmission in the case
$\pi \frac{n}{\SIZE} r^2 = (1+\ell) \ln{n}$.
\end{theorem}

\begin{proof} We refer to \cite{OUR-PAPER} where asymptotic
coverage as well as connectivity properties are treated in details
for the ranges of transmission considered in the theorem \ref{DEGREE}.
\end{proof}

\beginremark{3}
The theorem \ref{DEGREE} above remains valid for any bounded
surface $X$. Thus, it answers a conjecture of the authors of
\cite{PHILIPS-PANWAR-TANTAWI}.
\endremark

Next, we derive an upper-bound of the 
hop-diameter $\DIAM$ in the superconnectivity regime:

\smallskip

\begin{theorem}\label{DIAMETER}
Define by $\DIAM \equiv \DIAM(r)$ the hop-diameter of the graph
which is a priori a function of $r$ the transmission range.
$\DIAM$ is the maximum number of
hops required to travel from any node $u$ to another node $v$ of the
network. Suppose that
the transmission range satisfies 
$r = \sqrt{ \frac{(1+\ell)\ln{n}}{ \pi n} \, \SIZE }$ with
$\ell > 0$. We then have:
\begin{itemize}
\item[(i)] If $\ell > \frac{4-\pi}{\pi -2}$ then
\beq
\lim_{n \ten \infty} Pr \left[  \, D  \leq %
 3 \, \sqrt{\frac{\pi \, n}{(1+\ell) \, \ln n}}  + O(1) \, \right] = 1 \, .
\label{EQ:DIAM1}
\eeq
\item[(ii)] If $\ell \leq \frac{4-\pi}{\pi -2}$ then
\beq
\lim_{n \ten \infty} Pr \left[  \, D  \leq %
 5 \, \sqrt{\frac{\pi \, n}{(1+\ell) \, \ln n}}  + O(1) \, \right] = 1 \, .
\label{EQ:DIAM2}
\eeq
\end{itemize}
\end{theorem}

\begin{proof} Split the square $X$ into $j^2$ equal subsquares
$S_1, \, S_2, \, \cdots, S_{j^2}$. Each of the subsquares has a side
 $\sqrt{\SIZE}/j$ and an area $\SIZE/j^2$. Choose $j$ such
that each subsquare $S_i$ can entirely a circle of radius
equals to $r$ as depicted below.

\vspace{-0.0cm}
\begin{figure}[h]
  \begin{minipage}[t]{1.0cm}
    \begin{center}
      \begin{pspicture}(0,-1)(8,2)
        \psset{unit=0.75}
      	\pspolygon (2,0)(2,2)(0,2)(0,0)(2,0) 	
	\pscircle (1.0,1.0){1.0}
	\psline[linewidth=2pt]{<->}(0,1)(2,1)
	\rput(1,1.5) {$\frac{\sqrt{\SIZE}}{j}$}
	\psset{arrows=-}
	\pspolygon (3,0)(3,2)(5,2)(5,0)(3,0)
	\rput(3,3){{Subdivision of $X$}}
	\psline[linewidth=2pt]{->}(3,2.5)(4,1.5)
	\psset{arrows=-}
	\psline(3.2,0)(3.2,2)
	\psline(3.4,0)(3.4,2)
	\psline(3.6,0)(3.6,2)
	\psline(3.8,0)(3.8,2)
	\psline(4,0)(4,2)
	\psline(3,0.2)(5,0.2)
	\psline(3,0.4)(5,0.4)
	\psline(3,0.6)(5,0.6)
	\psline(3,0.8)(5,0.8)
	\psline(3,1)(5,1)
	\rput(4.5,1.3){$\mathbf{\cdots}$}
	\rput(1.5,-0.5){Size $\SIZE/j^2$}
	\psline[linewidth=0.75pt,linestyle=dotted]{->}(3,-0.5)(3.9,0.7)
      \end{pspicture}
    \end{center}
  \end{minipage}
  \hfill
  \begin{minipage}[t]{12.0cm}
    That is $\frac{\sqrt{\SIZE}}{2j} = r = \sqrt{\frac{(1+\ell)\ln{n} \SIZE}{\pi n}}$.
So, $j = \frac{1}{2} \sqrt{ \frac{\pi n}{(1+\ell) \ln{n}} }$. For sake of simplicity
but w.l.o.g., we suppose that $j \in \mathbb{N}$.
By 
the theorem \ref{DEGREE} (iii) given above, with high probability we have 
$\Theta(\ln{n})$ nodes inside the circle. Any pair of
processors inside the same circle need \textit{at most} $2$ hops to be connected
since they are at distance at most $2r$ and since each subgraph inside
such a circle is a.a.s. connected.
  \end{minipage}
  \hfill
\end{figure}
\vspace{0.1cm}

\noindent We claim that from two adjacent 
subsquares $S_1$ and $S_2$, communications
between any node $a \in S_1$ and any node $b \in S_2$ need at most
(w.h.p.)~: 
\begin{itemize}
\item[a)] $6$ hops for $\ell > \frac{4-\pi}{\pi -2} = 0.7519... $ and
\item[b)] $10$ hops for $\ell \leq \frac{4-\pi}{\pi -2}$.
\end{itemize}
To prove the first part, viz. a), consider adjacent subsquares as follows~: 
\vspace{-0.0cm}
\begin{figure}[h]
  \hfill
  \begin{minipage}[t]{5.0cm}    
    \begin{center}
      \begin{pspicture}(0,0)(4,4)
	\psset{unit=0.60}
	\rput(2.0,4.0){\small\textsc{At most $6$ hops}\normalsize}
	\rput(3.5,3.0){$L_1$}
        \psline[linewidth=1pt]{->}(3.5,2.8)(2.5,1.2)
	\rput(0.2,1.3){\BULLET}
	\rput(1.0,0.2){\BULLET}
        \psline[linewidth=0.25pt]{->}(0.2,1.3)(1.0,0.2)
	\rput(1.5,0.6){\BULLET}
        \psline[linewidth=0.25pt]{->}(1.0,0.2)(1.5,0.6)
	\rput(2.1,0.2){\BULLET}
        \psline[linewidth=0.25pt]{->}(1.5,0.6)(2.1,0.2)
	\rput(2.6,0.3){\BULLET}
        \psline[linewidth=0.25pt]{->}(2.1,0.2)(2.6,0.3)
	\rput(3.6,1.0){\BULLET}
        \psline[linewidth=1pt]{->}(2.6,0.3)(3.6,1.0)
	\rput(3.2,1.8){\BULLET}
        \psline[linewidth=1pt]{->}(3.6,1.0)(3.2,1.8)
	\psset{arrows=-}
      	\pspolygon[linewidth=2pt] (2,0)(2,2)(0,2)(0,0)(2,0) 	
      	\pspolygon[linewidth=2pt] (2,0)(2,2)(4,2)(4,0)(2,0) 	
	\pscircle[linestyle=dashed] (1.0,1.0){1.0}
	\pscircle[linestyle=dashed] (3.0,1.0){1.0}
	\pscircle[linestyle=dashed] (2.0,1.0){1.0}
      \end{pspicture}
      \label{FIG:HORIZONTAL}
  \vspace{0.22cm}
  \caption[]
  {Horizontal transmission.}
    \end{center}
  \end{minipage}
  \hfill
  \begin{minipage}[t]{5.0cm}    
    \begin{center}
      \begin{pspicture}(0,0)(4,4) 
	\psset{unit=0.60}
	\rput(1.4,3.6){$L_2$}
        \psline[linewidth=1pt]{->}(1.4,3.4)(2.5,2.5)
	\psset{arrows=-}
      	\pspolygon[linewidth=2pt] (2,0)(2,2)(0,2)(0,0)(2,0) 	
      	\pspolygon[linewidth=2pt] (2,4)(2,2)(4,2)(4,4)(2,4) 	
	\pscircle[linestyle=dashed] (1.0,1.0){1.0}
	\pscircle[linestyle=dashed] (3.0,3.0){1.0}
	\pscircle[linestyle=dashed] (2.0,2.0){1.0}
      \end{pspicture}
      \label{FIG:DIAGONAL}
    \end{center}
  \caption[]
  {Diagonal transmission.}
  \end{minipage}
  \hfill
  \hfill
  \begin{minipage}[t]{5.0cm}    
    \begin{center}
      \begin{pspicture}(0,0)(4,4) 
	\psset{unit=0.60}
	\rput(0.75,4.0){\small\textsc{At most}\normalsize}
	\rput(0.75,3.5){\small\textsc{$10$ hops}\normalsize}	
	\rput(0.2,1.3){\BULLET}
	\rput(1.0,0.2){\BULLET}
        \psline[linewidth=0.25pt]{->}(0.2,1.3)(1.0,0.2)
	\rput(1.5,0.6){\BULLET}
        \psline[linewidth=0.25pt]{->}(1.0,0.2)(1.5,0.6)
	\rput(2.1,0.2){\BULLET}
        \psline[linewidth=0.25pt]{->}(1.5,0.6)(2.1,0.2)
	\rput(2.6,0.3){\BULLET}
        \psline[linewidth=0.25pt]{->}(2.1,0.2)(2.6,0.3)
	\rput(3.6,1.0){\BULLET}
        \psline[linewidth=1pt]{->}(2.6,0.3)(3.6,1.0)
	\rput(3.0,1.2){\BULLET}
        \psline[linewidth=1pt]{->}(3.6,1.0)(3.0,1.2)
	\rput(3.8,2.2){\BULLET}
        \psline[linewidth=1pt]{->}(3.0,1.2)(3.8,2.2)
	\rput(2.7,2.6){\BULLET}	
        \psline[linewidth=1pt]{->}(3.8,2.2)(2.7,2.6)
	\rput(2.4,3.6){\BULLET}	
        \psline[linewidth=1pt]{->}(2.7,2.6)(2.4,3.6)
	\rput(3.9,3.9){\BULLET}	
        \psline[linewidth=1pt]{->}(2.4,3.6)(3.9,3.9)
	\psset{arrows=-}
      	\pspolygon[linewidth=2pt] (2,0)(2,2)(0,2)(0,0)(2,0) 	
      	\pspolygon[linewidth=2pt] (2,4)(2,2)(4,2)(4,4)(2,4) 	
      	\pspolygon[linewidth=2pt] (2,0)(2,2)(4,2)(4,0)(2,0) 	
	\pscircle[linestyle=dashed] (1.0,1.0){1.0}
	\pscircle[linestyle=dashed] (3.0,1.0){1.0}
	\pscircle[linestyle=dashed] (3.0,3.0){1.0}
	\pscircle[linestyle=dashed] (3.0,2.0){1.0}
	\pscircle[linestyle=dashed] (2.0,1.0){1.0}
      \end{pspicture}
      \label{FIG:DIAGONAL2}
    \end{center}
  \caption[]
  {``Undirect'' transmission.}
  \end{minipage}
\end{figure}

\noindent
A bit of trigonometry shows that 
 each lens-shaped region such as $L_1$ has a surface of
exactly  $|L_1| = \frac{1}{6}(4 \pi - 3 \sqrt{3}) r^2$. Note that
$L_1$ represents the intersection of two disks of equal
radius $r$ whose centers are at distance $r$. Therefore,
 there is no node inside the lens-shaped region $L_1$
 with probability 
\beq
\left(1- \frac{|L_1|}{\SIZE} \right)^n = %
\left( 1 -  \frac{1}{6}(4 \pi - 3 \sqrt{3}) \frac{(1+\ell) \ln{n}}{n} \right)^n %
 \leq \exp{\left( - \frac{1}{6}(4 \pi - 3 \sqrt{3}) (1+\ell) n \right)} \, . 
\eeq
Since each subsquare has at most $4$ lenses of size $|L_1|$, none
of these regions is empty with probability at least
\ben
\left(1- \exp{\left( - \frac{1}{6}(4 \pi -%
 3 \sqrt{3}) (1+\ell) n \right)}\right)^{(4 \times j^2)} & = & %
\left(1-  \frac{1}{n^{\frac{1}{6}(4 \pi -%
 3 \sqrt{3}) (1+\ell)}} \right)^{\frac{\pi n}{(1+\ell) \ln n}} \cr
& \geq & %
 \exp{\left(- 2 \times \frac{\pi n^{1-\frac{1}{6}(4 \pi -%
 3 \sqrt{3}) (1+\ell)}} {(1+\ell) \ln n} \right) }\, .
\label{hozizon} 
\een
Hence, with probability tending to $1$ as $n \ten \infty$, in every
lens-shaped region of size $|L_1|$ there is at least a node. Thus,
to transmit message between two horizontally (or vertically) adjacent subsquares,
we need at most 6 hops (see Figure 4).

To prove b), we consider lenses such as $L_2$ depicted in Figure 5.
The size of such region is
$|L_2| = \frac{ \pi-2 }{2} r^2$  which measures the area of the
 intersection of two equal disks
of radius $r$ and at distance $\sqrt{2} \times r$. Arguing as
for (\ref{hozizon}), we find that for every lens of size $|L_2|$ 
to be non-empty (w.h.p.) we need that $(1+\ell)\frac{\pi-2}{2} > 1$.
This condition is only satisfied if $\ell > \frac{2}{\pi - 2} - 1 = 0.7519...$.
For values of $\ell \leq 0.7519...$, transmissions are sent
horizontally then vertically (or vice-versa). Such transmissions 
can required up to $10$ hops (cf. Figure 6).
The proof of the theorem is now easily completed by simple counting arguments.
\end{proof}

\beginremark{4}
In theorem \ref{DIAMETER}, we consider transmitting ranges of
the form $\sqrt{ \frac{(1+\ell) \ln{n} }{ \pi n} \, \SIZE }$.
It is obvious that if the transmitting range augments,
 the hop-diameter diminishes. For instance, for values of transmitting
range satisfying  $\pi \frac{n}{\SIZE} r^2 = \omega(n) \,  \ln{n}$, with 
$\omega(n) \gg 1$, a.a.s. $\DIAM \leq 3 \, \sqrt{\frac{\, \pi \, n}{\ln n}}$. 
We refer the reader to 
the paper \cite{SHAKKOTTAI} where the authors
obtain similar results (with the notations of our paper, they
obtained  upper-bounds for $r > \sqrt{\frac{4 \ln{n} \SIZE}{\pi n}}$).
We also remark here that our theorem \ref{DEGREE} yields 
an immediate lower-bound of $\DIAM$ to finally show
that if $r$ satisfies  $r \geq (1+\ell) \, r_{\textsc{con}}$ for 
any $\ell > 0$ then  w.h.p.
 $\DIAM = \Theta(\sqrt{\frac{n}{\ln n}})$.
\endremark

\smallskip

\subsection{Adjusting the transmission range and fundamental limits}
\noindent
 The previous paragraph gives us almost sure characteristics of the network
but we need to verify and to exchange these informations by means of
 distributed protocols. To this end, we need two procedures. The first one
is the \proto{BROADCAST} protocol. In this protocol, some processors
(called \textit{sources}) try to diffuse a given message to all the
nodes in the network. It makes several calls of \proto{SEND}.
 The second procedure \proto{SFR} (for ``search-for-range'')
 is used to adjust the correct transmission range of the
 nodes in order to ``take control'' of the main 
 characteristics of the network. 
 It works as follows. 

Each processor starts with the maximum range of transmission. Then
 at each step, the transmission range is diminished gradually
until the deconnexion of some of the nodes. At this stage, these 
isolated nodes readjust their transmission range (in order to be
re-connected) and make call to \proto{BROADCAST}
 to send a message of ``deconnexion''
 to all the processors in the network. 
A processor quits the protocol
if and only if either it has been isolated once, is reconnected
and has sent the ``deconnexion'' message 
or it has received the ``deconnexion'' message containing
information about the adequate transmission range.

Now for details, we start with the \proto{BROADCAST} protocol.
The procedure is similar to the one in \cite{EXP-GAP} except that
we use \proto{SEND} to transmit messages.

\baselineskip=12pt

\noindent
\rule{\textwidth}{0.4pt}\\
\noindent {\textbf{Procedure} \proto{BROADCAST}($msg$, $\varepsilon$, $\Delta$, $r$, $N$)} \\
\noindent $k := 2 \lceil \log_2 \Delta \rceil$   \hfill %
($\star$ $\Delta$ is an upper-bound of the maximum degree  $\star$) \\
\noindent $\tau := \lceil \log_2\left( N/\varepsilon\right) \rceil$  \hfill %
($\star$ $N$ is an upper-bound of the number of nodes  $\star$) \\
\noindent Wait until receiving a message $msg$ \\
\noindent {\bf For} $i$ from $1$ to $\tau$ {\bf do} \\
\indent \indent \indent %
   Wait until $\mbox{\proto{TIME} mod } k = 1$  \hfill ($\star$ to synchronize $\star$) \\
\indent \indent \indent %
   \proto{SEND}$(msg,k,r)$                    \hfill ($\star$ attempt to send $msg$ $\star$) \\
\noindent {\textbf{end.}} \\
\rule{\textwidth}{0.4pt}

\baselineskip=16pt

\smallskip
\noindent
In the procedure above, $\varepsilon >0$ can be made arbitrarily small.
$\Delta$ is a parameter representing the maximum degree of the network
(or an upper bound of the maximum degree. This can be computed 
for a given value of 
the transmission range using theorem \ref{DEGREE}). $N$ is an upper-bound of the number
of participating nodes. \proto{TIME} is a protocol which allows a given node to have
the current time. Following the proves of \cite[Theorem 4]{EXP-GAP}, we have:

\smallskip

\begin{theorem} \label{BROADCASTTIME} \textbf{Bar-Yehuda, Goldreich, Itai \cite{EXP-GAP}.} 
Suppose that $r$ is the actual transmission range of the nodes.
Assume that $\Delta$ (resp. $N$) is an upper-bound of the maximum degree (resp.
the number of nodes) in the network and let
$T= 2\DIAM + 5\times \max{(\sqrt{\DIAM}, \sqrt{\log_2{(N/\varepsilon)}})} \times %
 \sqrt{\log_2{(N/\varepsilon)}}$. Assume that some initiators start 
the procedure \proto{BROADCAST}($msg$, $\varepsilon$, $\Delta$, $r$, $N$) at \proto{TIME} $=0$.
Then, with probability $\geq 1-2 \varepsilon$ by time $2 \lceil \log_2 \Delta \rceil T$,
all the nodes receive the message. Furthermore, with  probability $\geq 1-2 \varepsilon$
all the nodes have terminated by time 
$2 \lceil \log_2 \Delta \rceil ( T + \lceil \log_2 (N/\varepsilon) \rceil)$.
\end{theorem}

\beginremark{5} Note that in the procedure \proto{BROADCAST} above, we substitute
the \proto{DECAY} in \cite{EXP-GAP} by \proto{SEND} since our protocol 
\proto{SEND} seems to be more efficient than \proto{DECAY} 
(cf. theorem \ref{TH_SEND} in this paper and \cite[theorem 1]{EXP-GAP}). However, this
would only affect by a constant factor the time needed to accomplish a complete
broadcast in the network as given above.
\endremark

\noindent
We need to have as fast as
possible bounds of the value of the number $n$ of the processors.
If $p_0 = \lfloor \log_2 n \rfloor$ then $2^{p_0} \leq n < 2^{p_0+1}$.
Thus, by setting $R(2^p) := \sqrt{ \frac{ (\ln{(2^p)} + 2\ln{2})  \, \SIZE }{\pi 2^p} }$ 
 if the value of $p$ increases, $R(2^p)$ decreases. In the protocol \proto{SFR},
 we increment $p$ one by one, starting at a value close to the 
maximal transmission range of the processors. When $p$ passes through
$p_0-1$, $p_0$ and $p_0+1$, there will be w.h.p. some nodes which become 
isolated. In fact, a bit of calculus shows that
\ben
 \sqrt{ 2 \times \frac{\ln{n} \, \SIZE}{ \pi n}} \leq R(2^{p_0-1}) \, .
\een
We are now ready to give the protocol \proto{SFR}.
The procedure \proto{SFR} is executed by each station and the details follow~:

\newpage

\def\C1{\Big{\lceil} \frac{\ln \left( 1/2\, \varepsilon \right)}%
  {\ln \left( 2\, \varepsilon \right) }\Big{\rceil}}

\baselineskip=12pt

\noindent
\rule{\textwidth}{0.4pt}\\
( L0)  \noindent {\textbf{Procedure} \proto{SFR}($\varepsilon$)} %
\hfill {($\star$ ``Search-For-Range'' $\star$)} \\
( L1)  \noindent  \textbf{BEGIN} \\
( L2)  \indent $R := x \longmapsto  %
\sqrt{ \frac{ ( \ln{(2^x)} + 2\ln{2} ) \, \SIZE }{\pi \, 2^x} }$ ; %
\hfill {($\star$ Set $R$ as a {\bf local function}: $R \equiv R(x)$ $\star$)} \\
( L3)  \indent $B := x \longmapsto 24 %
 \Big{\lceil} \ln{x} \times \left( \sqrt{\frac{2^{x}}{x}} + x - \log_2{(\varepsilon}) \right) \Big{\rceil}$ ;
\hfill {($\star$ Similarly, the ``broadcast time'': %
 $B \equiv B(x)$ $\star$)} \\
( L4)  \indent ${\mbox{DECONNECTED}} := \mbox{ false}$ ; \hfill %
 {($\star$ Flag for isolated nodes $\star$)}\\
( L5)  \indent {$p := \Big{\lceil}\log_2{(r_\textsc{max})}\Big{\rceil}$ ; \hfill %
 {($\star$ Set the initial value of $p$ to the maximum $\star$)} \\
( L6)  \indent  \textbf{REPEAT} \\
( L7)  \indent \indent \indent $ counter := 0$;   \hfill {($\star$ counter for each station$\star$)} \\
( L8)  \indent \indent \indent $ t := max(4 \Big{\lceil} \log_2{(p)} \Big{\rceil}, %
                                       \Big{\lceil} \frac{\ln{(2/\varepsilon)}}{\ln 5} \Big{\rceil}) $; \\
( L9)  \indent \indent \indent \textbf{For} $i$ from $1$ to $t$ \textbf{Do} \\
(L10)  \indent \indent \indent \indent \indent     \proto{SEND}(``p'', $i$, $R(p)$); %
\hfill {($\star$ Succession of $i$ trials $\star$)} \\
(L11)  \indent \indent \indent \indent \indent     \textbf{If} receiving a message ``p'' \textbf{Then} \\
(L12)  \indent \indent \indent \indent \indent \indent \indent         $counter := counter+1$; \\
(L13)  \indent \indent \indent \indent \indent     \textbf{EndIf} \\
(L14)  \indent \indent \indent \textbf{EndFor} \\
(L15)  \indent \indent \indent \textbf{If} $counter = 0$ \textbf{Then} \hfill %
($\star$ The considered station received no message $\star$) \\
(L16)  \indent \indent \indent \indent \indent \textbf{For} $j$ from 1 to $\C1$ \textbf{Do} \\
(L17)  \indent \indent \indent \indent \indent \indent \indent     \proto{BROADCAST}(``Deconnexion $p$'',%
                  $\, \varepsilon$, $3 p$, $R(p-1)$, $2^{p+1}$) ; \\
(L18)  \indent \indent \indent \indent \indent \textbf{EndFor} \\
(L19)  \indent \indent \indent \indent \indent ${\mbox{DECONNECTED}} := \mbox{ true}$ ; \\
(L20)  \indent \indent \indent \textbf{Else} \\
(L21)  \indent \indent \indent \indent \indent Wait for a message for $\C1 \times B(p-1)$ times ; \\
\indent \indent \indent \indent \indent %
\indent \indent \indent \indent \indent %
 ($\star$ Give sufficient time to the advertisements of %
possible isolated stations $\star$) \\
(L22)  \indent \indent \indent \indent \indent \textbf{If} receiving the deconnexion message \textbf{Then} \\
(L23)  \indent \indent \indent \indent \indent \indent \indent Scan the value of $p$ and 
           set ${\mbox{DECONNECTED}} := \mbox{ true}$; \\
(L24)  \indent \indent \indent \indent \indent \textbf{Else}  $p := p+1$ ; \\
(L25)  \indent \indent \indent \indent \indent \textbf{EndIf} \\
(L26)  \indent \indent \indent \textbf{EndIf} \\ 
(L27)  \indent  \textbf{UNTIL} ${\mbox{DECONNECTED}} = \mbox{ true}$ ; \\
(L28)  \textbf{END.}\\
\rule{\textwidth}{0.4pt}\\

\baselineskip=16pt

When reaching the value of $p_0$, the isolated nodes -- 
whose transmission ranges are now set to $r= R(2^{p_0})$ -- can
increase back their transmission range, viz. $R(2^{p_0-1})$,
in order to be reconnected.  
Next, these processors have to advert the others about 
upper-bounds of $n$, $\Delta$ and $\DIAM$, respectively
given by 
\beq
n \leq 2^{p_0+1}, \quad \Delta \leq \frac{-1}{- W_0 (-e^{-1}/2)} %
\ln{n} < 3\, p_0%
\mbox{   and   } \DIAM \leq %
5 \sqrt{ \frac{\pi \, 2^{p_0}}{\ln{2} (p_0 +1)}} < %
12 \left[\sqrt{\frac{2^{p_0}}{p_0}}\right] \, ,
\label{UPPER_BOUNDS}
\eeq
where we used theorems \ref{DEGREE} and \ref{DIAMETER} for $\Delta$ and $\DIAM$.
The advertisements can be made correctly by means of multiple uses of the protocol \proto{BROADCAST}
but we have to give sufficient time slots -- cf. (L21) --
to the broadcasting processors in order to 
 let the others be aware of the bounds given by (\ref{UPPER_BOUNDS}). 
  The message sent for these advertisements is represented 
by a special message, 
say  ``\textit{Deconnexion} $p_0$'' which contains the right value of $p_0$.
Taking into account (\ref{UPPER_BOUNDS}),
we remark that the ``broadcast time'' given by the theorem \ref{BROADCASTTIME}
 is (with probability greater than $1-2\varepsilon$) 
less than $2 \lceil \log_2 \Delta \rceil  \times  %
(  2\DIAM + 5\times \max{(\sqrt{\DIAM}, \sqrt{\log_2{(N/\varepsilon)}})} \times %
 \sqrt{\log_2{(N/\varepsilon)}} + \lceil \log_2 (N/\varepsilon) \rceil)$. This
is strictly less than 
$24  \ln{(p_0)} \times \left( \sqrt{\frac{2^{p_0}}{p_0}} + p_0 - \log_2{(\varepsilon}) \right)$.
The protocol \proto{SFR} has the following properties:

\smallskip

\def\C2{ \frac{\ln \left( 1/2\, \varepsilon \right)}%
  {\ln \left( 2\, \varepsilon \right) } }

\begin{theorem} \label{TH:SFR}
Suppose that the random deployed network is an instance that 
has exactly the upper-bounds given by (\ref{UPPER_BOUNDS}), that
is the main characteristics $n$, $\Delta$ and $\DIAM$ 
of the input graph satisfy 
(\ref{UPPER_BOUNDS}) with probability $1$.
For any $c>0$ there exist a constant $c_1 > 0$ such that
 with probability at least $1 - \frac{1}{n^{c}}$,
 the protocol \proto{SFR}($\frac{1}{n^{c_1}}$) terminates 
in at most 
$O( \ln{\ln{n}}  \, \sqrt{ \frac{n}{\ln{n}}} \,)$ 
time slots. After this time,
with probability at least $1-\frac{1}{n^{c}}$, every node is aware of
the upper-bounds of values of $n$, $\Delta$ and $\DIAM$. 
\end{theorem}

\begin{proof} Due to place limitation, we will just discuss
the main lines of the proof. In lines (L9)--(L14), the inner 
loop is repeated $t$ times. Consider a random picked node $v$.
By theorem \ref{TH_SEND}, as soon as
 $i$ (cf. line (L9)) satisfies $2^i \gg d_v$ ($d_v$ represents
the degree of $v$), the probability of success of each call
of \proto{SEND} is at least $0.8...$. By theorem \ref{DEGREE},
and under the hypothesis that the graph satisfies the almost
sure properties of a random network, 
if $p = p_0 = \lfloor  \log_2{(n)} \rfloor$, $d_v < 3 p_0$.
Therefore, by setting $t$ as in line (L8), we insure that 
if the node $v$ is still connected, it will receive more than
one  message from its neighbors with probability 
at least $1-\frac{\varepsilon}{3}$.  
Similarly, by repeating sufficient calls of \proto{BROADCAST}
for the just deconnected nodes (see the discussion above)
 and give sufficient time to them to send the
message of deconnexion to the others, we give sufficient chance to the
processors of the whole network 
to know the correct upper-bounds of $n$ (and thus $\Delta$
and $\DIAM$). To explain the constants $c$ and $c_1$ involved in
the result, one can always choose $\varepsilon$ of the form
$\varepsilon = 1/n^{c_1}$ in order to obtain probabilities of
failure of order $1/n^{c}$.
\end{proof}

Despite theorem \ref{TH:SFR} is given with the assumption that
the input network satisfies (\ref{UPPER_BOUNDS}), we believe that
 \proto{SFR} with slight modifications can handle the other cases
(which indeed occur extremely rarely). For instance, the line
(L17) can be modified in this sense and one can give the upper-bounds
in (\ref{UPPER_BOUNDS}) greater values (e.g. by augmenting the actual 
values).

\section{The initialization protocol}
We have settle in the previous paragraph the problem of determining
the correct transmission range for the nodes of a random network
 to have the characteristics (mainly maximum degree and
hop-diameter) dicted by theorems \ref{DEGREE} and
\ref{DIAMETER}. We also know through the protocol \proto{SFR}
a probabilistic upper-bound of the number of participating
nodes. In \cite{EMULATION}, Bar-Yehuda \textit{et al.} gave
 protocols for efficient emulation of
 a single-hop network with collision detection on multi-hop radio network,
provided that the number of nodes, the diameter and the maximum degree
of the network (or upper-bounds of them) are known. Combination
of their results with ours lead to a new initialization protocol.
In the next paragraph, we will first give the protocol for the
single-hop network without collision detection
and then extend it to our purpose by means of the emulation protocols
given in \cite{EMULATION}.

\smallskip

\subsection{Initialization in the single-hop model.}
In the single-hop model of networks \cite{OLARIU}, we have
direct links between any pair of processors.
First, we give a procedure that can randomly split a set 
$S$ of directly connected processors, into many subsets,
say $S_1$, $S_2$, $\cdots$, $S_k$, such that at least
two of the subsets are non-empty. In this subsection, we
consider that the processor has the collision-detection ability.
This assumption doesn't change our final results 
 thanks to the emulation protocols given in \cite{EMULATION}.
The following procedure \proto{EQUIPARTITION} is an implementation
of Bernoulli process in order to partition a given initial set
into $k$ subsets.  The process is repeated until the original set
is partionned into at least two nonnull subsets. Note that, the ``non-empty status''
can be checked by the nodes since it happens if and only if
 either there is collision between two or more messages or 
there is exactly one station that has put a message on the channel.

\beginremark{6} It is important to remark here that these protocols
are originally due to Hayashi, Nakano and Olariu \cite{HAYASHI} (see also \cite{OLARIU}) but
we put them here for sake of completeness.
\endremark

\baselineskip=12pt

\noindent
\rule{\textwidth}{0.4pt}\\
\noindent \textbf{\textsc{Procedure}} \proto{EQUIPARTITION}(\textbf{Input:} $S$, $k$
, \textbf{Var:} $\mbox{\texttt{Number}}$;
\textbf{Output:} ${S}_1$, ${S}_2$, $\cdots$, ${S}_k$) \\
 \indent \indent \textbf{Repeat} \\
 \indent \indent \indent \indent each station selects randomly an integer $i \in \{1, 2, \cdots, k\}$ \\
 \indent \indent \indent \indent to join $S_i$ and broadcasts when $\mbox{\proto{TIME} mod } k = i$ ; \\
 \indent \indent \textbf{Until} \textit{at least} two of the subsets %
$S_i, i \in \left[1, k\right]$ are \textbf{non-empty} \\
 \indent \indent \textbf{For} $i$ from $1$ to $k$ \textbf{Do} \\
 \indent \indent \indent \indent \textbf{If} $|S_i| = 1$ \textbf{Then} \\
 \indent \indent \indent \indent \indent \indent \indent \indent the unique station is labeled with $N$ %
and quits the protocol; \\
 \indent \indent \indent \indent \indent \indent \indent \indent $\mbox{\texttt{Number}} := \mbox{\texttt{Number}}+1$; \\
 \indent \indent \indent \indent  \textbf{EndIf} \\
 \indent \indent \textbf{EndFor} \\
 \textbf{\textsc{END.}} \\
\rule{\textwidth}{0.4pt}

\baselineskip=16pt

\noindent 
Therefore, in \proto{EQUIPARTITION} the test ``at least two of the subsets %
$S_i, i \in \left[1, k\right]$ are \textbf{non-empty}'' for the repeat loop
can easily done in the single-hop case with collision detection (in fact,
 stations can distinguish between the lack of message, the presence
of exactly one message and collisions). Consider now a group of $m$
stations, the probability of failure of the splitting process applied
on this subset is $Pr\left[\textit{failure}\right] = k^{-m+1}$. So, with
probability $1 -  k^{-m+1}$, \proto{EQUIPARTITION} subdivides the inital 
set of $m$ stations into at least $2$ subsets. For $i \in \left[1,k\right]$,
if $|S_i| = 1$, that is a processor is alone in a subset, it can be labeled.
The labeling of such processor is carried out with the variable 
$\mbox{\texttt{Number}}$
that is incremented each time a station leaves the protocol. The procedure
\proto{INITIALIZATION} is then called. The details are given as follows:

\baselineskip=12pt

\noindent
\rule{\textwidth}{0.4pt}\\
\noindent \textbf{Procedure} \proto{INITIALIZATION}(\textbf{Input}: $S$, $k$)\\
\noindent For each station, set $\mbox{\texttt{Number}} := 1$ and $L := 1$; \\
\noindent Initialize $P_L := \{$ all stations $\}$; \\
 \indent \textbf{While} $L \geq 1$ \textbf{Do} \\
 \indent \indent \indent \textbf{Switch} $|P_L|$ \\
 \indent \indent \indent \indent \indent  \textbf{Case} 0: \textbf{Case} 1: \\
 \indent \indent \indent \indent \indent  \indent \indent $L \leftarrow L-1$; \\
 \indent \indent \indent \indent \indent  \textbf{Default}: \\
 \indent \indent \indent \indent \indent  \indent \indent Partition $P_L$ into 
 $P_L$, $P_{L+1}$, ..., $P_{L+k-1}$ \\
 \indent \indent \indent \indent \indent  \indent \indent%
 with \proto{EQUIPARTITION}($P_L$, $\mbox{\texttt{Number}}$, $k$)}; \\
 \indent \indent \indent \indent \indent  \indent \indent%
 $L := L+k-1$ ; \\
 \indent \textbf{EndWhile} \\
 \textbf{End.} \\
\rule{\textwidth}{0.4pt}

\baselineskip=16pt
The protocol \proto{INITIALIZATION} has the following property:
\begin{theorem} \label{TH_INITIALIZATION}
Let $S$ be a group of $n$ nodes communicating directly via a single channel. \\
\proto{INITIALIZATION}$(S,3)$ always terminates (with probability $1$)
and its running time is in average approximately $\frac{3 }{\ln{3}} n$ time slots.
\end{theorem}

\begin{proof} The proof of this theorem relies on
Mellin transform asymptotics \cite{KNUTH,PHILIPPE-ROBERT} and is
omitted. We refer also to the papers \cite{HAYASHI,OLARIU} for 
randomized protocols in the single-hop cases and to \cite{MYOUPO}
for the average-case analysis of such algorithms.
\end{proof}

\subsection{Initialization of the wireless multihop network}
\noindent
We turn now on a more general self-configuration protocol designed for 
wireless multihop network. In the light of the previous paragraphs,
one can design an initialization protocol as follows~:
\begin{itemize}
\item[\textbf{Step 1)}] Use \proto{SFR} for determine upper-bounds of
$n$, $\Delta$ and $\DIAM$,
\item[\textbf{Step 2)}] Emulate the protocol \proto{INITIALIZATION}
for single-hop networks described above in order to 
 give IDs for the nodes of the wireless multihop network.
To this purpose, a natural idea is to repeat the number of tests
to check how many stations are actually broadcasting.
This can be done by means of the protocol that allows the
emulation of collision detection given in \cite{EMULATION}. By theorem
\ref{TH_INITIALIZATION} and since each broadcast
costs $O(\sqrt{n \ln n})$ time slots, the initialization of
a multihop wireless network can be done in subquadratic
time slots, viz.  $O( \sqrt{n^{3} \ln{n}})$.
With extremely small probability, there will be duplicate IDs. 
These errors are rare and can be checked once
the nodes are ``identified'' with the help of the
 \textit{deterministic} distributed 
$O(n^{3/2} \ln^2{n})$ gossiping protocol given
in \cite{CHROBAK}.
\item[\textbf{Step 3)}] In the presence of failures,
reiterate the process by augmenting the values
given in \textbf{Step 1)} and go directly to \textbf{Step 2)}.
All together, combinations of these algorithms
lead to an initialization  protocol which 
always terminates in expected time $O(n^{3/2} \ln^2{n})$. 
\end{itemize}

\section{Conclusion}
We showed that given a randomly distributed wireless nodes with
density $n/\SIZE$, when the transmission range of the nodes 
 is set to $r = \sqrt{(1+\ell) \frac{\ln n \, \SIZE}{\pi \, n}}$~: 
(i) the hop-diameter is less than $5 \sqrt{ \frac{\pi n}{(1+\ell) \ln{n}}}$,
 (ii) the network is $\Theta(\ln{n})$-connected,
each point of the support area is monitored by $\Theta(\ln{n})$ nodes and 
the degrees of all nodes are $\Theta(\ln{n})$, with high probability.
We showed how these results can help
to conduct precise analysis in order to design
protocols for the self-configuration of the network. The protocols
of this paper are fully distributed and assume only as \textit{a priori}
knowledge of the nodes the size of the support area $X$. These
results illustrate how fundamental limits of 
random networks can help researchers and
developpers for the design of algorithms in the extremal scenarios and
the protocols given in this paper can serve as basis for other
 decentralized algorithms.

As a final comment, it is important to note that for the single-hop
cases, \textit{energy-efficient} protocols have been designed by
Nakano and Olariu \cite{OLARIU2}.
Their works naturally suggest a generalization for 
the multihop cases under various scenarios.

\newpage

\bibliographystyle{plain}

\newpage

\section*{Appendix}
\begin{figure}[h]
\hfill
  \begin{minipage}[t]{6.5cm}
    \begin{center} 
      \psfig{file=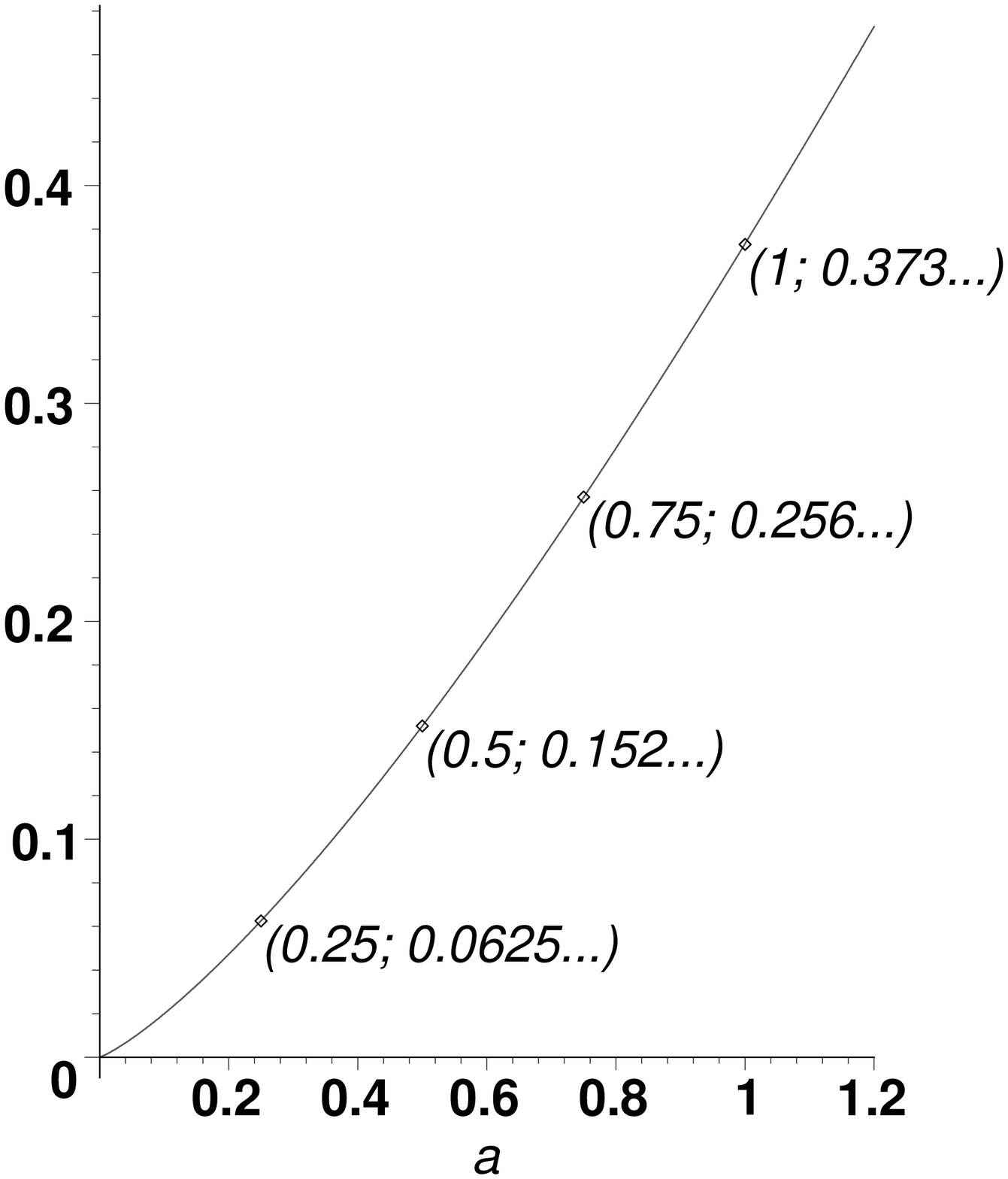,width=6.2cm,height=6.2cm}
    \end{center}
    \caption[A plot of $ -\ell/W_{-1}\L(-\frac{\ell}{e\, (1+\ell)}\R)$.]
    {A plot of $ -\ell/W_{-1}\L(-\frac{\ell}{e\, (1+\ell)}\R)$.}
    \label{FIG:PLOTW1}
  \end{minipage}
\hfill
  \begin{minipage}[t]{6.5cm}
    \begin{center}
      \psfig{file=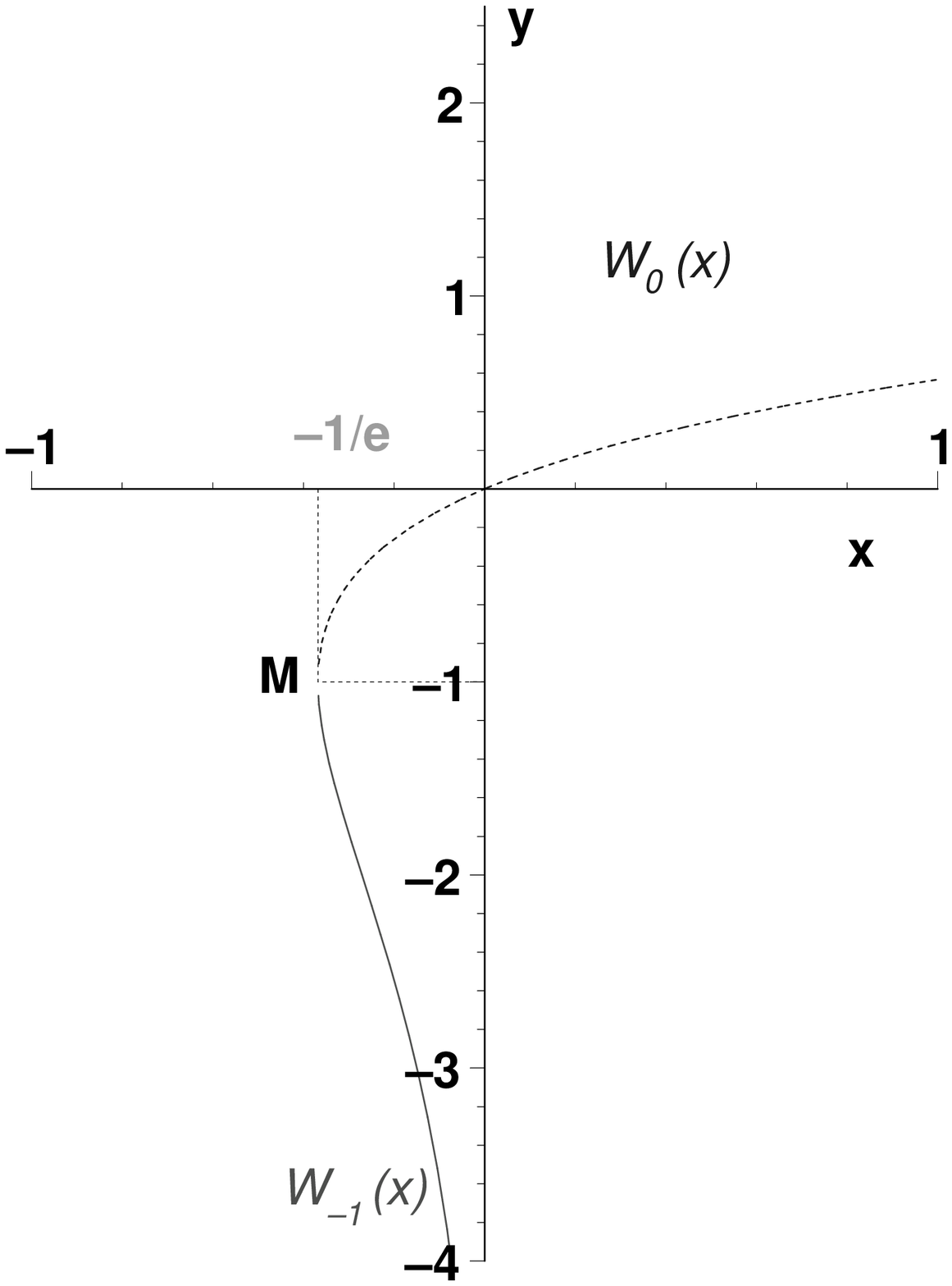,width=6.2cm,height=6.2cm}
    \end{center}
    \caption{The branches $W_0$ (dashed line) and $W_{-1}$ 
      (solid line) of the Lambert W function.}
    \label{TWO-BRANCHES}
    \end{minipage}
\hfill
\end{figure}

\noindent {\bf The Lambert W function.} In this paragraph,
we give some properties of the function satisfying $W(x)e^{W(x)} = x$.
We remark here that the function $W$, in particular
the principal branch $W_0$, already plays
a central key role when studying the random graph model
$\G(n,m)$, i.e., the random graph built with $n$ vertices and $m$ edges
 which is the ``enumerative counterpart'' of
the $\G(n,p)$ random graph model (see,  e.g., 
the ``giant paper'' \cite{JKLP93}).
In fact, $-W_0(-x)$ is the exponential generating function that enumerates
 Cayley's rooted trees \cite{CAYLEY} and we have
\beq
-W_0(-x) = \sum_{i=1}^{\infty} \frac{n^{n-1} x^n}{n!} \, .
\label{EGF-CAYLEY}
\eeq
We plot in Figure \ref{TWO-BRANCHES} the two real branches of 
the Lambert W function considered in this paper.
This function has been recognized
 as solutions of many problems in various fields of mathematics,
physics and engineering as emphasized in \cite{LAMBERTW}.
The Lambert W is considered as a special function of
mathematics on its own and its computation has been implemented
in mathematical software as Maple. Figure \ref{TWO-BRANCHES}
represents the two real branches of the Lambert W function.
It is shown that the two branches meet at point $M=(-1/e,\, -1)$.
As an example, if $\ell = \frac{1}{2}$ in (\ref{DILBERT}),
each point of the area $X$ is covered, with high probability,
 by at least  $\sim .1520088850 \ln{n}$ disks. We have the Figure
\ref{FIG:PLOTW1} depicting the function 
$\ell \ten -\ell/W_{-1}\L(-\frac{\ell}{e\, (1+\ell)}\R)$
involved in eq. (\ref{DILBERT}).

\vspace{-0.0cm}
\vspace{0.75cm}

\end{document}